# Coherence Resonance in Models of an Excitable Neuron with Both Fast and Slow Dynamics

Robert C. Hilborn[i] and Rebecca J. Erwin[ii]

*Department of Physics, Amherst College, Amherst, Massachusetts 01002*

We demonstrate the existence of noise-induced periodicity (coherence resonance) in both a discrete-time model and a continuous-time model of an excitable neuron. In particular, we show that the effects of noise added to the fast and slow dynamics of the models are dramatically different. A Fokker-Planck analysis gives a quantitative explanation of the effects.

PACS number(s): 05.40.-a, 05.45.-a, 87.17.Nn

The effects of noise on nonlinear dynamical systems are often counter-intuitive. Under appropriate conditions, noise can enhance the response of a nonlinear system to an external signal (stochastic resonance) [1]. For other systems, noise can enhance the periodicity of an oscillatory component of the system's behavior, an effect known as autonomous stochastic resonance [2, 3], coherence resonance [4] or stochastic coherence [5]. Coherence resonance has been demonstrated in models of single excitable systems [3, 4, 6-11], coupled excitable systems [12-14], chaotic systems [15], coupled chaotic systems [16, 17], spatio-temporal arrays [18, 19] and a few experimental systems [13, 15, 20-24].

In this Letter, we focus on coherence resonance in models of an excitable neuron. Real neurons often display (at least) two time scales: fast dynamics corresponding to action potentials and slow dynamics corresponding to chemical concentration variations. We show that coherence resonance is dramatically different depending upon whether the noise is added to the fast dynamics or to the slow dynamics. This is a new feature of coherence resonance and enriches the spectrum of noise-induced phenomena. The effect should also be important in the interpretation of experimental observations of coherence resonance in any system with multiple time scales.

We first consider a discrete-time (iterated map) model [25] whose behavior mimics that of physiological neurons [26, 27]. The model has two dynamical variables: one corresponding to the membrane voltage in a neuron and the other to a gating-ion concentration (usually $Ca^{2+}$ in actual neurons). We demonstrate that adding noise to the voltage variable produces a coherence resonance effect quite different from that obtained when noise is added to the gating-ion concentration.

The iterated map model, augmented with additive noise terms, is given by a set of coupled, discrete-time functions for the dynamical variables $x_n$ and $y_n$:

$$x_{n+1} = \frac{\alpha}{1 + x_n^2} + y_n + D_x \xi_{xn} \tag{1}$$

$$y_{n+1} = y_n - \beta x_n - \sigma + D_y \xi_{yn} . \tag{2}$$

The subscript $n$ indicates the iteration number. $D\xi$ is the external noise term, which we take to be a white-noise, Gaussian-distributed source with zero mean and variance $D^2$. Noise in $x$ represents, for example, synaptic input noise in the neuron membrane voltage; noise in $y$ models ion-concentration fluctuations, which may be either external to the cell or internal [28]. The difference in behavior as a function of $D_x$ as compared to $D_y$ is the main subject of this Letter. The parameters $\alpha$, $\beta$, and $\sigma$ set the operating conditions for the model. We use $\beta = \sigma = 0.001$ as in Ref. [25]. Since these parameters are small compared to 1, the $y$ time dependence is slow compared to that of $x$.

Rulkov [25, 26] and de Vries [27] studied the behavior of Eqs. (1)-(2) with $\alpha$ in the range 4.0-4.9. Here we focus on the parameter range near $\alpha = 2$. For $\alpha$ just greater than 2, the system exhibits periodic pulses. For $\alpha < 2$, the behavior, after transients die out, approaches a stable fixed point at $x = -1$, $y = -1 - \alpha/2$. For $\alpha < 2$, noise added to either $x$ or $y$ (or both) will induce pulses if the noise fluctuation is sufficiently large to move the system outside the basin of attraction of the (no-noise) fixed point. Figure 1 illustrates the behavior of the iterates of Eqs. (1)-(2) for $\alpha = 1.99$, $D_x = 0.005$, and $D_y = 0$. We see that $x$ exhibits intervals of pulse behavior followed by periods of quiescent behavior, and thus $x$ is analogous to the membrane voltage in physiological neurons. The $y$ variable oscillates slowly over a smaller range of values and is analogous to the relatively slowly changing chemical concentrations in neurons.

We now turn to the main point of this Letter: the effect of noise on the periodicity of the system. As a measure of the periodicity of the system's behavior, we calculate the "regularity" $R$ defined as [29, 30]

$$R = \frac{\langle T \rangle}{\sqrt{\operatorname{var}(T)}}, \qquad (3)$$

where $\langle T \rangle$ is the mean time between pulses and $\operatorname{var}(T)$ is its variance. Figure 2 shows $R$ as a function of noise amplitude $D$ for two cases: one with $D_y = 0$ and the other with $D_x = 0$. Each data point represents an average over five independent noise realizations of 100,000 iterations of Eqs. (1)-(2). Both cases show a clear coherence resonance effect: a maximum in the regularity as a function of noise amplitude. Figure 3 shows $R$ as a function of both $D_x$ and $D_y$. As $D_y$ increases, the maximum of $R$, as a function of $D_x$ decreases rapidly.

In the excitable regime, the behavior can be divided into two parts [4, 9, 10, 31, 32]: (a) an activation phase that lasts until the noise is able to move the system away from the (no-noise) fixed point and (b) a phase for the large excursion through state space. For small noise values, the excursion time is essentially unaffected by noise, while the activation time decreases as the noise amplitude increases, leading to an increase in $R$. For large noise amplitudes, noise moves the system away from the fixed point as soon as the excursion is over, and the excursion time itself begins to fluctuate because of the noise, leading to a decrease in $R$. The two opposing tendencies yield a maximum in $R$ as a function of noise amplitude. For small values of the noise amplitude, a Poisson distribution describes the distribution of the time intervals, and $R \to 1$.

For the case of noise added to the $y$ (slow) variable, Fig. 2 shows that the maximum $R$ occurs at a smaller noise amplitude than for the case of noise added to the $x$ (fast) variable. Furthermore, the maximum value for the regularity is considerably smaller [10, 13]. The physical explanation of these features has two parts: (a) The size of the fixed point's basin of attaction in the $y$ direction is about 0.016 the size in the $x$ direction. (When noise kicks the system outside this basin of attraction, a pulse occurs.) Thus, the relevant noise amplitude regime for noise added to $y$ is approximately 0.016 of that for noise added to $x$ as seen in Fig. 2. (b) The maximum in $R$ occurs when the noise has reduced the activation time to a value smaller than the excursion time. The effects of noise on the excursion time then determine the maximum value of $R$. As we discuss below, the behavior of $y$ during an excursion is more easily disrupted by noise than is the corresponding $x$ behavior. Thus, the maximum regularity for noise added to the slow variable is less than that observed for noise added to the fast variable.

The activation phase can be described as a first-passage (or first-exit) time problem, where the mean first-passage time and its variance can be calculated from the Fokker-Planck equation describing the time-evolution of the probability distribution for the system [33-35]. For a one-dimensional system in the interval $[b,a]$, the first and second moments of the first-passage time are given by

$$\langle T(w) \rangle = \frac{2}{D^2} \int_w^a dz \int_b^z du\, e^{2(U(z)-U(u))/D^2} \tag{4}$$

$$\langle T^2(w) \rangle = \frac{4}{D^2} \int_w^a dz \int_b^z du\, e^{2(U(z)-U(u))/D^2} T(u)\ , \tag{5}$$

where $U$ is the function whose gradient gives the deterministic part of the dynamics, and $w$ is the so-called injection point (initial location). The angular brackets indicate an average over noise realizations, $a$ is an absorbing boundary, and $b$ is a reflecting boundary. Although we are dealing with a discrete-time system, we may use the results from a continuous-time analysis because the changes in $x$ and $y$ per time step are relatively small. When noise is added to $x$ or $y$ alone, we can treat the behavior as approximately one-dimensional. Near the fixed point, the potential $U$ is determined from by linearizing Eqs. (1) and (2). Matching the measured mean activation time as a function of noise amplitude $D$ sets $a$, $b$, and $w$.

The excursion phase itself can be divided into two parts: a pulse time $T_p$ while $x \approx 0$, and a recovery time $T_r$ during which $x$ increases from $-1.5$ to $-1.0$. To account quantitatively for the noise-enhanced regularity, we need models for the pulse and recovery phases. For noise added only to $x$, the system behavior during the pulse and recovery segments essentially tracks the location of the fixed point of Eq. (1) (without the noise term) with $y$ viewed as a slowly changing parameter. In this approximation, the mean pulse and recovery times are independent of the noise, while the variance of these times is given by the behavior in a quadratic potential function with a time-dependent fixed-point location:

$$\mathrm{var}(T) = D^2 / (2kv^2). \qquad (6)$$

Here $k$ is the potential parameter near the (moving) fixed point and $v$ is the speed of the moving fixed point. The details of the analysis will be presented elsewhere [36].

When the noise is added to $y$ only, we treat the pulse and recovery phases as segments during which $y$ varies (approximately) linearly with iteration number as seen in Fig. 1. Since there is no "restoring force" in this case, the $y$ behavior is more easily disturbed by noise than is the corresponding fixed point motion for $x$. In both cases, the regularity of the behavior is then evaluated from the generalization of Eq. (3)

$$R = \frac{\langle T_a \rangle + \langle T_r \rangle + \langle T_p \rangle}{\sqrt{\mathrm{var}(T_a) + \mathrm{var}(T_r) + \mathrm{var}(T_p)}}. \qquad (7)$$

Figure 2 shows the results of this analysis for $\alpha = 1.99$. The agreement with the results of the numerical simulations is quite good given the simplicity of the approximations. Our approximations are expected to break down, however, for the larger noise amplitudes [36].

Neuron behavior has traditionally been described by differential equation models. To illustrate the fast/slow noise effect in continuous-time models, we use a version of the Morris-Lecar model [37] that has both fast and slow dynamics [38]. The model includes the neuron membrane voltage $v(t)$, a potassium-channel gating variable $w(t)$ and the (slowly varying) calcium-ion concentrations $[Ca](t)$ with noise terms added to the voltage and calcium-ion equations:

$$\dot{v}(t) = (-I_{ion}(v,w) + I_{ext})/C + \sqrt{D_v}\xi_v \qquad (8)$$

$$\dot{w}(t) = \phi[f(v) - w(t)]/\tau(v) \qquad (9)$$

$$[\dot{Ca}](t) = \varepsilon(\mu I_{Ca}(v) - [Ca]) + \sqrt{D_{Ca}}\xi_{Ca} \ . \qquad (10)$$

$I_{ion}$ is the total ion current through the membrane, determined by voltage-dependent conductances and ion concentrations. $I_{ext}$ is the externally controlled membrane current, here used as a control parameter. $C$ is the membrane capacitance per unit area. $\phi$ is a temperature-dependent factor, and $\tau(v)$ is a voltage-dependent time constant for the gating variable. $\varepsilon$ is a small parameter that sets the $[Ca]$ time scale relative to the voltage time scale. $\mu$ is a factor that converts the calcium current $I_{Ca}$ to a rate of change of concentration. The details of the model can be found in [38]. $\sqrt{D}\xi$ is a Gaussian-distributed random process with zero mean and variance $D$.

We use the parameter values given by Rinzel and Ermentrout [38] with the following changes: $\varepsilon = 0.0005$, $C = 17\,\mu\text{F/cm}^2$, and $I_{ext} = 44.173\,\mu\text{A/cm}^2$ to put the Morris-Lecar model in an excitable regime with one neuron spike per pulse and a large enough difference in time scales to show the desired effects. Eqs. (8)-(10) are integrated using the modified Euler method as described in [39]. Figure 4 shows the regularity as a function of noise amplitude for the Morris-Lecar model. The distinction between noise added to the fast (voltage) variable and noise added to the slow ($[Ca]$) variable is evident. A Fokker-Planck analysis of these results and a more detailed biological interpretation of these effects will be presented elsewhere [36].

In summary, we have demonstrated distinct fast/slow coherence resonance effects in both discrete-time and continuous-time models of an excitable neuron. A Fokker-Planck analysis gives a quantitative explanation of these differences for the discrete-time model. The small noise-levels required for coherence resonance in these models suggest that these effects may be important in physiological neurons. The voltage fluctuations are associated with fluctuations due to the large number of synaptic connections to a neuron. The gating-ion fluctuations are more difficult to measure and to control and may be due to extracellular concentration fluctuations or to fluctuations of the intracellular concentrations due to clustering of internal release mechanisms [28]. We expect to see the same fast/slow effect in other excitable systems with multiple time scales.

[i] Email address: rchilborn@amherst.edu

[ii] Current address: Department of Physics, California Institute of Technology

**Figure Captions**:

FIG. 1. The time dependence of the fast ($x$) and slow ($y$) variables of a Rulkov model oscillator described by Eqs. (1) and (2) for $\alpha = 1.99$, $\beta = \sigma = 0.001$, $D_x = 0.005$ and $D_y = 0$.

FIG. 2. A plot of the regularity $R$ of the Rulkov model as a function of noise amplitude with $\alpha = 1.99$. The solid squares are for noise added only to the $x$ variable. The solid circles are for noise added only to the $y$ variable. The vertical bars indicate the one-standard-deviation uncertainty from five noise realizations. For low and high values of the noise amplitude, the uncertainty is small. The solid curves are the results of the Fokker-Planck analysis described in the text.

FIG. 3. The regularity $R$ for the Rulkov model Eqs. (1) and (2) as a function of $\log_{10} D_x$ and $\log_{10} D_y$.

FIG. 4. The regularity $R$ for the Morris-Lecar model Eqs. (8)-(10) as a function of the noise amplitude $D_\upsilon$ with $D_{Ca} = 0$ (circles) and as a function $D_{Ca}$ with $D_\upsilon = 0$ (squares). The uncertainty bars indicate one-standard-deviation for five noise realizations.

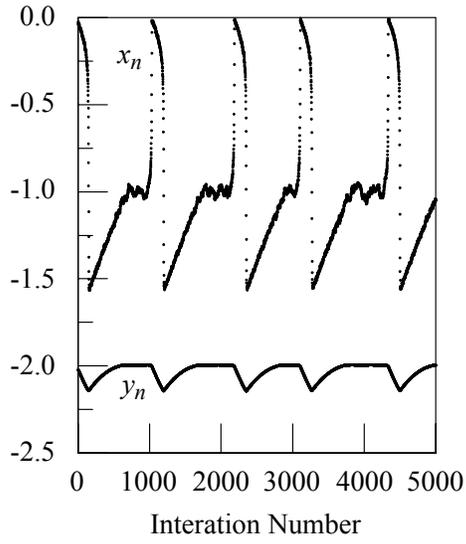

FIG. 1. The time dependence of the fast (*x*) and slow (*y*) variables of a Rulkov model oscillator described by Eqs. (1) and (2) for $\alpha = 1.99$, $\beta = \sigma = 0.001$, $D_x = 0.005$ and $D_y = 0$.

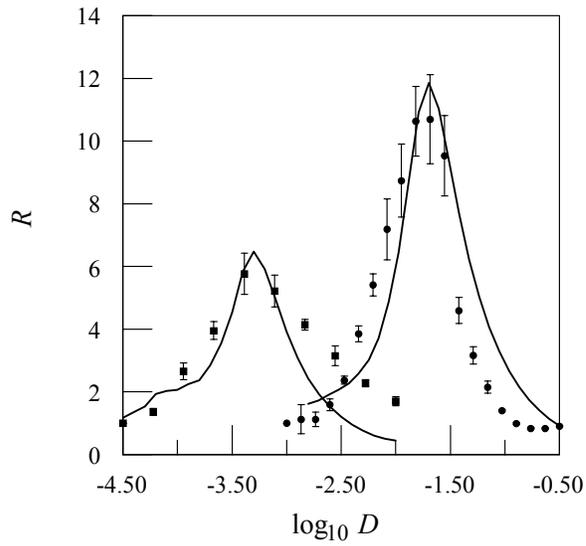

FIG. 2. A plot of the regularity $R$ of the Rulkov model as a function of noise amplitude with $\alpha = 1.99$. The solid circles are for noise added only to the $x$ variable. The solid squares are for noise added only to the $y$ variable. The vertical bars indicate the one-standard-deviation uncertainty from five noise realizations. For low and high values of the noise amplitude, the uncertainty is small. The solid curves are the results of the Fokker-Planck analysis described in the text.

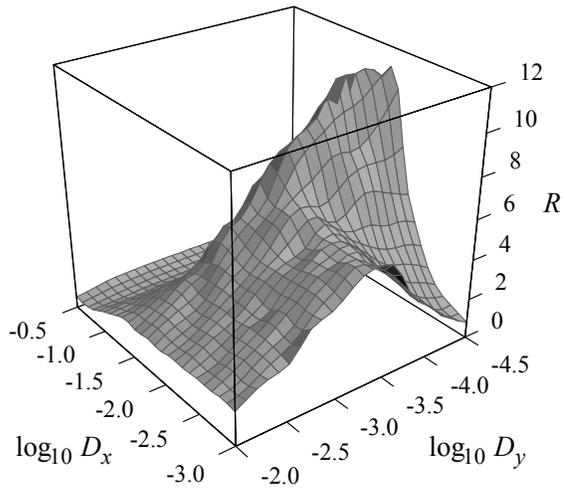

FIG. 3. The regularity $R$ for the Rulkov model Eqs. (1) and (2) as a function of $\log_{10} D_x$ and $\log_{10} D_y$.

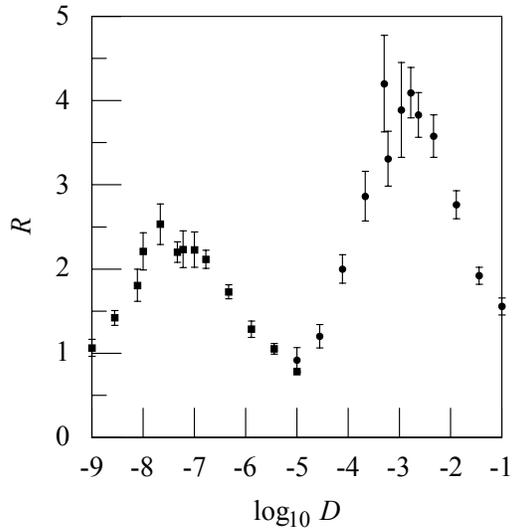

FIG. 4. The regularity $R$ for the Morris-Lecar model Eqs. (8)-(10) as a function of the noise amplitude $D_v$ with $D_{Ca} = 0$ (circles) and as a function $D_{Ca}$ with $D_v = 0$ (squares). The uncertainty bars indicate one-standard-deviation for five noise realizations.

---

[i] Email address: rchilborn@amherst.edu

[ii] Current address: Department of Physics, California Institute of Technology